\documentclass[10pt,letterpaper]{article}
\usepackage{opex3}
\usepackage{color}

\begin{document}

\title{Demonstration of a highly-sensitive tunable beam displacer with no movable elements based on the concept of weak value amplification}

\author{Luis Jos\'{e} Salazar-Serrano$^{1,2}$, David Guzm\'{a}n$^{2}$, Alejandra Valencia$^{2}$ and Juan P. Torres$^{1,3}$}

\address{$^1$ICFO-Institut de Ciencies Fotoniques, Mediterranean Technology Park, 08860 Castelldefels (Barcelona), Spain\\
$^2$Quantum Optics Laboratory, Universidad de los Andes, AA 4976, Bogot\'{a}, Colombia\\
$^3$Universitat Polit\`{e}cnica de Catalunya, Dept. of Signal
Theory \& Communications, 08034 Barcelona, Spain}

\email{$^*$luis-jose.salazar@icfo.es} %

\begin{abstract}
We report the implementation of a highly sensitive beam displacer
based on the concept of weak value amplification that allows to
displace the centroid of a Gaussian beam a distance much smaller
than its beam width without the need to use movable optical
elements. The beam's centroid position can be displaced by
controlling the linear polarization of the output beam, and the
dependence between the centroid's position and the angle of
polarization is linear.
\end{abstract}

\section{Introduction}
A polarization beam displacer (BD) is a device that splits an input polarized beam into two spatially separated beams that propagate parallel with orthogonal polarizations. Commercially available BD are made of birefringent materials like Calcite crystal, Barium Borate ($\mathrm{\alpha-BBO}$) crystal, Rutile crystal or Yttrium Vanadate ($\mathrm{YVO_4}$) among others. In these devices, due to the intrinsic birefringence of the material, the propagation direction of the ordinary polarized beam is unchanged whereas the extraordinary component deviates inside the crystal~\cite{Fowles1975}. The beam separation is fixed and its maximum value depends on the crystal material and length.

A BD can also be used to displace spatially the position of a single optical beam, for example by using an input beam with vertical polarization at the input. However, in many applications is desired to move the position of a single beam over a given interval~\cite{Fowler1966}. To the best of our knowledge, a scan of the position of a single beam can be implemented either by using an arrange of moving mirrors~\cite{Katz1995,Galvez2001}, a plane-parallel plate or a tunable beam displacer (TBD)~\cite{salazarRevSciInst2014}.

In the first case, a set of mirrors are arranged in a configuration that allows to change the position of the output beam when one or various mirrors are rotated. In the second case, a transparent plane-parallel plate of certain thickness such as a tweaker plate~\cite{Thorlabs}, a thin film polarizer~\cite{II-IV} or a plate beam splitter~\cite{Edmund} is rotated with respect to an axis parallel to the surfaces offsets the position of the input beam after consecutive refractions in the air-plate and plate-air interfaces. The beam displacement is proportional to the plate thickness and the rotation angle. Finally, in a TBD, two mirrors fixed to a platform are rotated with respect to a polarizing beam splitter (PBS). When the angle is different from zero, the input beam splits into two parallel propagating beams with orthogonal polarizations separated by a distance proportional to the rotation angle. If the input beam polarization is horizontal or vertical, a single beam is obtained at the output.

For all the cases mentioned above the beam shift results from the mechanical rotation of an optical element. This condition imposes a technical limitation on the sensitivity of the beam displacer since it directly relates to which sensitivity we can achieve when performing the rotation. In a plane-parallel plate displacer one can obtain a typical beam shift of $\approx 12.5\,\mathrm{\mu m/deg}$, where the proportionality factor depends on the thickness of the plate and its index of refraction. For a TBD, the proportionality factor is $\approx 5\,\mathrm{mm/deg}$ which depends mainly on the distance from the mirrors to the PBS.

In this paper we demonstrate an optical device that can outperform the limitations imposed by the use of movable optical elements. In our scheme, we do not make use of the tunable reflections or/and refractions induced by the rotation of a specific optical element. Instead, we make use of the concept of weak value amplification~\cite{aharonov1988,duck1989}, that allows to convert two beams with orthogonal polarizations that slightly overlap in space into a single beam whose center can be tuned by only modifying the linear polarization of the output beam.

\begin{figure}[t!]
\centerline{\includegraphics[height=0.25\textheight]{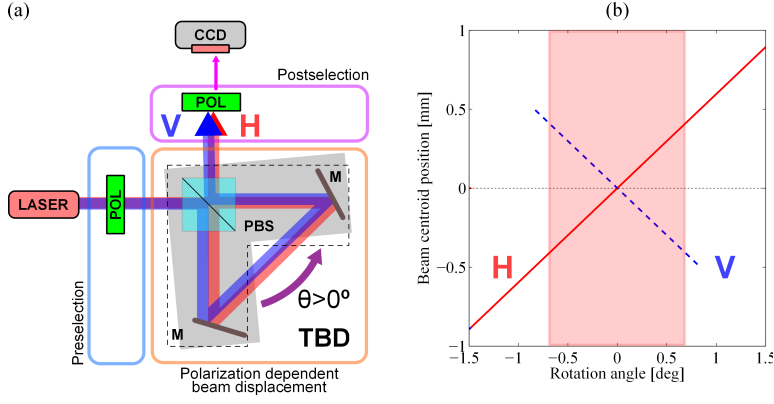}}
\caption{(a) General scheme of the tunable beam displacer. A
polarization-dependent beam displacement is introduced by rotating
the polarizing beam splitter (PBS) an angle $\theta$ with respect
to the direction of propagation of the input Gaussian beam. Input
and output polarizers (POL) control the corresponding
polarizations. (b) Beam displacement before traversing the second
polarizer for the horizontal (solid line) and vertical (dashed
line) components of the optical beam a function of the rotation
angle $\theta$. The shaded region indicates the region where the
beams with orthogonal polarizations still overlap.}
\label{fig:figure1}
\end{figure}

\section{Scheme for a highly sensitive tunable beam displacer}
Fig. \ref{fig:figure1} (a) shows the general scheme of the beam
displacer. It is based on the device described by Feldman et al.
\cite{Feldman2006} with the difference that our device does not
use quarter waveplates that limit the spatial quality of the beam
and the wavelength range of operation.  A laser generates an input
Gaussian beam with amplitude $E_{\mathrm{in}}(x,y)=E_0
\exp\left[-(x^2+y^2)/(2w^2)\right]$, where $E_0$ is the peak
amplitude, and $w$ is the $1/e$ beam width. The polarization of
the input beam is selected to be $\textbf{e}_{\mathrm{in}} =
(\textbf{x}+\textbf{y})/\sqrt{2}$, with the help of a polarizer. A
polarizing beam splitter (PBS), rotated a small angle $\theta$
with respect the direction of propagation of the input beam,
splits the input beam into two output beams with orthogonal
polarizations, where the horizontal component is shifted a small
distance $+\Delta x$ with respect to the input beam centroid,
while the vertical component is shifted a distance $-\Delta x$.
Fig. \ref{fig:figure1} (b) shows the beam centroid displacement
for each polarization as a function of the TBD rotation angle
($\theta$). The TBD is set to operate in the shaded region shown
in Fig. \ref{fig:figure1} (b), where the two output beams with
orthogonal polarizations still overlap. i.e., the distance between
the two beam centroids ($2\Delta x$) is small compared to the beam
diameter ($w$).

After recombination of the two orthogonal beams, slightly
displaced one with respect to the other a distance $2\Delta x$,
and projection into the polarization state
$\textbf{e}_{\mathrm{out}} =
\cos\beta\,\textbf{x}+\sin\beta\,\textbf{y}$ by using a second
polarizer, the amplitude of the output beam writes
\begin{eqnarray}
& & {\bf{E}}_{\mathrm{out}}(x,y) = \frac{E_0\cos\beta}{\sqrt{2}}
\exp\left\{[-\left[(x-\Delta
x)^2+y^2\right]/(2w^2)+i\,\phi\right\} \nonumber
\\
& & +\frac{E_0\sin\beta}{\sqrt{2}} \exp\left\{-\left[(x+\Delta
x)^2+y^2\right]/(2w^2)\right\}\,, \label{eq:Eout}
\end{eqnarray}
where $\phi$ takes into account any optical path difference
between the orthogonal polarizations that could have been
introduced, i.e., due to misalignment between the optical beams
that leaves the PBS through different output ports.

\begin{figure}[t!]
\centerline{\includegraphics[height=0.25\textheight]{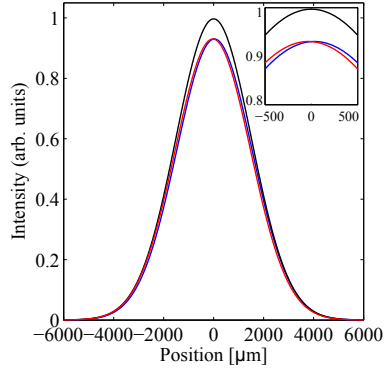}}
\caption{Beam profile after traversing the second polarizer for
three different output polarizations ($\beta=30^{\circ}$,
$\beta=45^{\circ}$ and $\beta=60^{\circ}$). The insets shows more
clearly the small beam displacements for different post-selections
of the output state of polarization.}
\label{fig:figure2}
\end{figure}

Since the spatial shape of the beam in the $x$ and $y$ directions
are independent, and the displacement is only considered along the
$x$ direction, for the sake of simplicity we will be looking only
at the beam shape along the $x$ direction. The intensity of the
output beam, $I_{\mathrm{out}}(x)=|E_{\mathrm{out}}(x)|^2$ writes
\begin{eqnarray}
I_{\mathrm{out}}(x) &=&
\nonumber\frac{I_0}{2}\Big\{\cos^2\beta\exp\left[-(x-\Delta
x)^2/w^2\right]+\sin^2\beta\exp\left[-(x+\Delta x)^2/w^2\right]
\nonumber \\
& & +\exp\left(-\Delta
x^2/w^2\right)\sin2\beta\exp\left(-x^2/w^2\right)\cos\phi\Big\}\,.
\label{eq:Iout}
\end{eqnarray}
We fix the angle $\theta$, which generates a certain displacement
$\Delta x$, as shown in Fig. \ref{fig:figure1}(b). Fig.
\ref{fig:figure2} shows the output intensity, after traversing the
second polarizer, for three different angles: $\beta=30^{\circ}$,
$\beta=45^{\circ}$ and $\beta=60^{\circ}$. An angle
$\beta=45^{\circ}$ corresponds to choosing the polarization of the
output beam equal to the polarization of the input beam.
Inspection of Fig. \ref{fig:figure2} shows that
$I_{\mathrm{out}}(x)$ corresponds to a single peaked Gaussian-like
distribution whose center is slightly shifted with respect to the
input beam centroid by an amount smaller than $\Delta x$, far less
than the beam width. We also observe that this small shift is
polarization-dependent, i.e., it depends on the value of the angle
$\beta$. This effect can be easily visualized by calculating the
beam's centroid $\langle x \rangle= \int
x\,I_{out}(x)\,\mathrm{d}x / \int I_{out}(x)\,\mathrm{d}x$. We
also show the insertion loss (expressed in decibels)
$L=-10\log_{10}[P_{\mathrm{out}} / P_{\mathrm{in}}]$ where
$P_\mathrm{in}$ and $P_\mathrm{out}$ designate the input and
output power of the beams, respectively. The
polarization-dependent shift is always associated with a similarly
polarization-dependent insertion loss.

\begin{figure}[t!]
\centerline{\includegraphics[height=0.25\textheight]{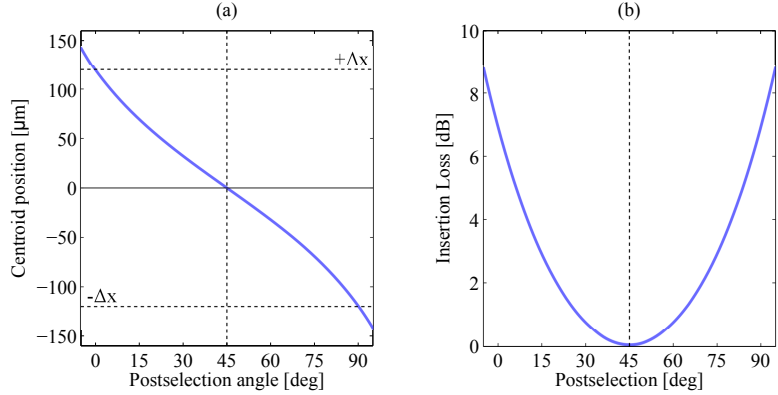}}
\caption{(a) Centroid position as a function of the polarization
selected of the output beam, given by the post-selection angle
$\beta$. (b) Insertion loss as a function of the post-selection
angle $\beta$. Data: $\Delta x = 120\,\mathrm{\mu m}$,
$\gamma=0.9$ and $\phi=0^{\circ}$.} \label{fig:figure3}
\end{figure}

Making use of Eq. (\ref{eq:Iout}), the centroid of the output beam
can be written as
\begin{equation}
\langle x \rangle =
\frac{\cos2\beta}{1+\gamma\sin2\beta\cos\phi}\,\Delta x\,.
\label{eq:centroid}
\end{equation}
where $\gamma=\exp\left(-\Delta x^2/w^2\right)$ is close to unity
since $\Delta x \ll w$. Similarly, the insertion losses is given
by
\begin{equation}
L =
-10\log_{10}\left[\frac{1}{2}\left(1+\gamma\sin2\beta\cos\phi\right)\right]\,.
\label{eq:losses}
\end{equation}
Figs. \ref{fig:figure3} (a) and Fig. \ref{fig:figure3} (b) show
the beam centroid position and the insertion loss as a function of
the output polarizer angle (post-selection angle $\beta$). The
displacements $\pm \Delta x$ for each polarization are indicated
by horizontal dashed lines.

Equation (\ref{eq:centroid}) shows that the beam centroid $\langle
x \rangle$ is related to the polarization-dependent displacement
$\Delta x$ by a relationship of the form $\langle x \rangle =
A\cdot \Delta x$, where $A=\cos2\beta
\left[1+\gamma\sin2\beta\cos\phi\right]^{-1}$ is the amplification
factor. Most applications of the weak value amplification concept
(see, for instance, \cite{dressel2014} and \cite{Jordan2014} for
two recent reviews about this topic) are interested in a regime
where $A \gg 1$. However this is not the only regime where weak
value amplification can be of interest ~\cite{Torres2012}. Here,
on the contrary, we are interested in the regime $A \ll 1$, where
beam displacements much smaller than the beam width of the input beam are observed.
In this regime, close to
$\beta=45^{\circ}$ (input and output polarizations are similar)
the centroid position of the output beam varies almost linearly
with respect to the postselection angle over the range $-\Delta x
\leq \langle x \rangle \leq +\Delta x$ [see Fig. \ref{fig:figure3}
(a)], and the insertion loss is small for the same interval [see
Fig. \ref{fig:figure3} (b)], making the weak value amplification
scheme described in Fig. \ref{fig:figure1} (a) suitable for
implementing a low-loss highly sensitive tunable beam displacer
where the spatial shift is controlled by projection into a given
polarization state, with no movable optical elements.

\section{Experimental demonstration}
In order to demonstrate the feasibility of the tunable beam
displacer discussed above, we implement the set-up shown in Fig.
\ref{fig:figure1}(a). The input beam is a He-Ne laser (Thorlabs
$\mathrm{HRP005S}$) and the input beam is Gaussian with a beam
waist of $\sim 600\,\mathrm{\mu m}$ ($1/e^2$). Two Glan-Thomson
polarizers (Melles Griot $\mathrm{03PT0101/C}$) are used to select
the initial and final states of polarization before and after the
TBD. The initial state of polarization is selected by rotating the
first polarizer at $+45^{\circ}$, and the output polarization is
selected by rotating the second polarizer an angle $\beta$ with
respect to the horizontal direction.

The TBD is composed of two aluminum mirrors, positioned
equidistantly from a $1.0\,\mathrm{cm}$ polarizing beam splitter
(PBS), and fixed to a L-shaped platform that is free to rotate an
angle $\theta$ with respect to the PBS center. For a given angle,
the separation between the two output beams depends on $\theta$,
the distance from the mirrors to the PBS, and the sizes of the
input beam and the PBS. In the setup, the distance from each
mirror to the PBS is set to $7\,\mathrm{cm}$ and the platform is
rotated with a motorized rotation stage.

The output beam cross section is detected by a CCD camera (Santa
Barbara Instruments ST-1603ME) with $1530\times1020$ pixels
($9\,\mathrm{\mu m}$ pixel size). With the data measured, the
corresponding centroid position is calculated using a simple
MATLAB program. To avoid CCD saturation, neutral density
absorptive filters (Thorlabs - Serie NE-A) are used.

Before running the experiment an initial alignment is carried out
without using the output polarizer. This preparation consists of
two steps. Firstly, the input beam enters the TBD, $\theta$ is set
to zero and the angle for each mirror is set such that each beam
reflected on the mirrors propagates towards the PBS center and
only one beam is seen in the camera. The centroid of this image
sets the reference point from which the new beam's centroid
position, $\langle x \rangle$, will be measured. Secondly, the
L-shaped plaque is rotated by an angle $\theta$ to define the
small initial displacement, $\Delta x$, between the components
with orthogonal polarization. For our experiment, $\Delta
x=120\,\mathrm{\mu m}$, which yields $\gamma=\exp(-\Delta
x^2/w^2)$ equal to $0.96$. Once the reference centroid is defined,
the output polarizer is introduced. A set of images are recorded
for different values of $\beta$, and their corresponding centroids
are calculated.

\begin{figure}[t!]
\centerline{\includegraphics[height=0.25\textheight]{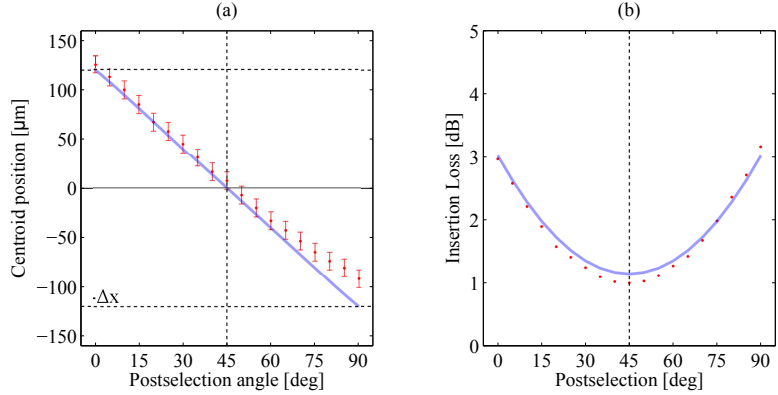}}
\caption{Panel (a), measurement (dots) of the beam's centroid
position (left axis), amplification factor $A$ (right axis), and
fit (solid line) following Eq. (\ref{eq:centroid}) as a function
of the postselection angle $\beta$. Panel (b), measured insertion
loss (dots) and fit (solid line) obtained from Eq.
(\ref{eq:losses}) as a function of $\beta$.} \label{fig:figure4}
\end{figure}

The experimental results are presented as dots in Fig.
\ref{fig:figure4}. Panel (a) depicts the measured beam
displacement $\langle x\rangle$ as a function of the output polarizer angle ($\beta$).
The error bars take into account the uncertainty introduced by the
CCD camera pixel size of $9\,\mathrm{\mu m}$. The solid line in
Fig. \ref{fig:figure4} (a) corresponds to the best data fit using
Eq.(\ref{eq:centroid}) where $\phi$ is the fitting parameter. From
the best fit we obtain $\phi=54^{\circ}$, which corresponds to a
difference in optical path of $\sim 0.094\,\mu m$, mainly due to
misalignment. In the region $0^{\circ}\leq\beta\leq 90^{\circ}$ we
observe that the beam's centroid varies almost linearly with
respect to the output polarizer angle. In this interval, the best
fit gives $\langle x\rangle = -2.32\,\beta +114.24\,\,\mu m$,
which demonstrates a region of operation that goes approximately
between $-120\,\mathrm{\mu m}$ to $+120\,\mathrm{\mu m}$,  in
agreement with the initial displacement of $\Delta
x=120\,\mathrm{\mu m}$. The sensitivity of the shift is limited by
the angular resolution achievable when selecting the output
polarization. As an example, if a manual rotation mount with
resolution of $10\,\mathrm{arcmin}$ is used to select the output
polarization, a minimum beam displacement step of $380\,nm$ can be
obtained without using opto-mechanical components. In panel (b) we
show the measured (dots) and theoretical (solid line) insertion
loss, given by Eq. (\ref{eq:losses}) for $\phi=54^{\circ}$ and
$\gamma=0.96$. The maximum insertion loss in this region is $\sim
3\,\mathrm{dB}$.

\section{Conclusions}
In conclusion, we have implemented and demonstrated a low-loss
tunable beam displacer based on the concept of weak value
amplification that allows to displace the centroid of a beam with
very high sensitivity. Interestingly, the relationship between the
beam's centroid shift and the output polarization is is almost
linear, and the sensitivity of the beam displacement is limited by
the sensitivity available for selecting the output polarization.
From the measurements, we were able to shift the centroid of a
Gaussian beam with a beam waist of $\sim 600\,\mathrm{\mu m}$,
over an approximate interval that goes from $-120\,\mathrm{\mu m}$
to $+120\,\mathrm{\mu m}$ in steps of less than $ \sim 1 \mu m$.

\vspace{1cm} \noindent {\bf Acknowledgements}: JPT and LJSS
acknowledge support from the Spanish government (Severo Ochoa
programs), and from Fundaci\'o Privada Cellex, Barcelona. LJSS and
AV acknowledges support from Facultad de Ciencias, U. de Los
Andes. LJSS would like to thank Luis Carlos G\'{o}mez from the
mechanical workshop in U. de Los Andes for his valuable help in
implementing the custom made rotating stage used to rotate the
output polarizer. AV also acknowledges support from Vicerrector\'{i}a
de Investigaciones and FAPA project at U. de Los Andes.


\begin{thebibliography}{99}
\bibitem{Thorlabs} For example, the tweaker plate from Thorlabs model XYT-A is a $2.5\,\mathrm{mm}$ thick plane-parallel plate that allows sub-mm level precision beam displacement.
\bibitem{II-IV} II-VI UK LTD offers thin film polarizers made of either ZnSe or Ge that can be used to split or combine an input beam into two components with orthogonal polarizations. The polarizer is oriented at Brewster's angle with respect to the input beam so that the vertical polarization is highly reflected whereas the horizontal is transmited.
\bibitem{Edmund} For instance, Edmund optics plate beam splitter model \#49-684 is a $3\,\mathrm{mm}$ thick N-BK7 splitter that transmits $70\%$ of the input power and operates in the visible regime.
\end{thebibliography}
\end{document}